\documentclass[
aps,%
12pt,%
final,%
notitlepage,%
oneside,%
onecolumn,%
nobibnotes,%
nofootinbib,%
superscriptaddress,%
noshowpacs,%
centertags]%
{revtex4}
\usepackage{float}

\begin{document}

\title{Reconstruction the scalar-torsion gravity version\\
from the frame of exact cosmological solutions}

\author{\firstname{I.~V.}~\surname{Fomin}}
\email{ingvor@inbox.ru}
\affiliation{%
Bauman Moscow State University, Moscow, Russia
}%
\author{\firstname{S.~V.}~\surname{Chervon}}
\email{chervon.sergey@gmail.com}
\affiliation{%
Bauman Moscow State University, Moscow, Russia
}%
\affiliation{%
Ulyanovsk State Pedagogical University, Ulyanovsk, Russia
}%
\affiliation{%
Kazan Federal University, Kazan, Russia
}%
\author{\firstname{E.~S.}~\surname{Dentsel}}
\email{edentsel@yandex.ru}
\affiliation{%
Bauman Moscow State University, Moscow, Russia
}%


\begin{abstract}
We consider cosmological models based on the scalar-torsion gravity implying non-minimal coupling between torsion and the scalar field with certain relations between model's parameters. Based on observational constraints on the values of the parameters of cosmological perturbations, the type of the coupling was determined. It was noted that any inflationary models constructed on the basis of the proposed approach can be verified by observational constraints.
\end{abstract}

\maketitle

\section{Introduction}

Currently, a description of the inflationary stage of the evolution of the universe with including some inflaton fields for the case of general relativity and its various modifications is necessary for constructing correct cosmological models~\cite{Baumann:2014nda,Chervon:2019sey,Odintsov:2023weg}. At the moment there are a large number of cosmological models describing both stage of accelerated expansion of the universe with different types of the scalar fields based on the Einstein gravity theory~\cite{Baumann:2014nda,Chervon:2019sey}, the teleparallel equivalent of the general relativity (TEGR)~\cite{Aldrovandi:2013wha,Bahamonde:2021gfp} and their various extensions and modifications~\cite{Cai:2015emx,Fomin:2020hfh,Fomin:2022ozv,Gonzalez-Espinoza:2019ajd,Gonzalez-Espinoza:2020azh,Chervon:2023gio} are actively used.

For cosmological models based on general relativity (or TEGR), agreement with observational data will lead to an unambiguous determination of the model parameters including the potential of a scalar field~\cite{Baumann:2014nda,Chervon:2019sey,Bahamonde:2021gfp}.
However, for the case of cosmological models based on modified theories of gravity, agreement with observational data can be obtained for ambiguously defined parameters~\cite{Cai:2015emx,Fomin:2020hfh,Fomin:2022ozv,Gonzalez-Espinoza:2019ajd,Gonzalez-Espinoza:2020azh,Chervon:2023gio}.

For example, for the case of cosmological models based on the scalar-tensor gravity theories in paper~\cite{Fomin:2022ozv} it was shown that these models can satisfy observational constraints not due to the choice of model parameters, but due to a specific relationship between the parameters for arbitrary cosmological models.
To obtain such relation, the exact solutions of the cosmological dynamics equations for known types of scalar-tensor gravity  were used~\cite{Fomin:2022ozv}.

In this paper we consider the application of this approach to the case of cosmological models based on the scalar-torsion gravity with non-minimal coupling of the scalar field and torsion, which generalize the case of the teleparallel equivalent of general relativity.

The properties of the cosmological models with scalar-torsion gravity expressed in a fairly general form were considered earlier in~\cite{Gonzalez-Espinoza:2019ajd,Gonzalez-Espinoza:2020azh} based on the slow-roll approximation. Also, specific cosmological models based on the exact solutions of dynamic equations were considered in~\cite{Chervon:2023gio}.

In this case, we consider generalized exact solutions of the cosmological dynamic equations for inflation based on the scalar-torsion gravity for arbitrary Hubble parameter to reconstruct a class of verifiable inflationary models. Also, the relationship between the potential of the scalar field and the type of the scalar-torsion gravity was reconstructed in explicit form.

\section{Cosmological models in generalized scalar-torsion gravity}\label{A}

The generalized action for cosmological models based on the scalar-torsion gravity with non-minimal coupling of a scalar field $\phi$ and torsion $T$ in the system of units $c=8\pi G=M^{-2}_{p}=1$ can be written as~\cite{Gonzalez-Espinoza:2019ajd,Gonzalez-Espinoza:2020azh}
\begin{eqnarray}\label{act-1}
 S=\int d^{4}x\,e\, F(T,X,\phi)=\int d^{4}x\,e\,\left[f(T,\phi)+ \omega(\phi)X \right],
\label{action1}
\end{eqnarray}
where $f=f(T,\phi)$ is an arbitrary differentiable function of a scalar field $\phi$ and torsion scalar $T$, also kinetic energy of a scalar field $X=-\partial_{\mu}{\phi}\partial^{\mu}{\phi}/2$.

The action (\ref{act-1}) includes non-minimally coupled scalar-torsion gravity models with $f(T,\phi)$
and the kinetic function $\omega=\omega(\phi)$.

In choosing the cosmological background, we assume the diagonal tetrad field
\begin{equation}
\label{veirbFRW}
e^A_{~\mu}={\rm diag}(1,a,a,a),
\end{equation} which is a proper tetrad naturally associated with the vanishing spin connections $\omega^{A}_{~ B\mu}=0$, and which leads to the spatially flat Friedmann-Robertson-Walker (FRW) metric~\cite{Gonzalez-Espinoza:2019ajd,Gonzalez-Espinoza:2020azh}
\begin{equation}
ds^2=-dt^2+a^2\,\delta_{ij} dx^i dx^j \,,
\label{FRWMetric}
\end{equation} where $a=a(t)$ is the scale factor and $t$ is the cosmic time.

The background equations for this model can be written as~\cite{Gonzalez-Espinoza:2019ajd,Gonzalez-Espinoza:2020azh}
\begin{eqnarray}
\label{00}
f(T,\phi) - \frac{1}{2}\omega(\phi)\dot{\phi}^{2} - 2Tf_{,T} =0, \\
\label{ii}
f(T,\phi) + \frac{1}{2}\omega(\phi)\dot{\phi}^{2}
- 2 T f_{,T} - 4\frac{d}{dt}\left(Hf_{,T}\right)=0, \\
\label{phi}
\omega(\phi)\ddot{\phi}+3\omega(\phi)H\dot{\phi}
+\frac{1}{2}\frac{d\omega(\phi)}{d\phi}\dot{\phi}^{2}-f_{,\phi}=0,
\end{eqnarray}
where $H\equiv \dot{a}/a$ is the Hubble rate, and a dot represents derivative with respect to $t$.
Partial derivatives is denoted as $f_{,\phi}=\frac{\partial f}{\partial \phi},~
f_{,T}=\frac{\partial f}{\partial T}$.

It should be noted, that the torsion scalar is $T=6H^2$ for the case of the spatially flat Friedmann-Robertson-Walker metric~\cite{Gonzalez-Espinoza:2019ajd,Gonzalez-Espinoza:2020azh}.

From here, using the relation $H=\sqrt{\frac{T}{6}}$ and taking into account the condition $H>0$, we can rewrite the dynamic equations (\ref{00})--(\ref{phi}) as
\begin{eqnarray}
\label{A1}
&&f(T,\phi) - \frac{1}{2}\omega(\phi)\dot{\phi}^{2} - 2Tf_{,T} =0, \\
\label{B1}
&&\frac{1}{2}\omega(\phi)\dot{\phi}^{2}-2\frac{d}{dt}\left(\sqrt{\frac{T}{6}}f_{,T}\right)=0, \\
\label{C1}
&& \omega(\phi)\ddot{\phi}+3\omega(\phi)\sqrt{\frac{T}{6}}\dot{\phi}
+\frac{1}{2}\frac{d\omega(\phi)}{d\phi}\dot{\phi}^{2}-f_{,\phi}=0.
\end{eqnarray}

Such representation of dynamic equations allow us select the class of models, admitting exact solutions
of the equations (\ref{A1})--(\ref{C1}) for arbitrary Hubble parameter $H=H(t)$.

The exact solutions of the system of equations (\ref{A1})--(\ref{C1}) for the special form of the function $f(T,\phi)$ can be written as follow
\begin{eqnarray}
\label{SOL3}
&&f(T,\phi)=\alpha_{1} G(\phi)\sqrt{T}+\alpha_{2} V(\phi)=f(T,\phi)_{STG}+\alpha_{2}V(\phi),\\
\label{SOL4}
&&\omega(\phi)=\frac{\alpha^{2}_{1}}{3\alpha_{2}}\frac{G_{,\phi}^{2}}{V(\phi)},\\
\label{SOL4B}
&&\dot{G}=\dot{\phi}\,G_{,\phi}=\frac{\sqrt{6}\alpha_{2}}{\alpha_{1}}V(\phi),
\end{eqnarray}
where function $f(T,\phi)_{STG}=\alpha_{1}G(\phi)\sqrt{T}$ fixed the version of the scalar-torsion gravity, $G=G(\phi)$ is arbitrary differentiable function of a scalar field, and $\alpha_{1}$, $\alpha_{2}$ are the normalization constants.

Using expression $\ddot{\phi}=\frac{d\dot{\phi}}{d\phi}\dot{\phi}$, one can verify by direct substitution, that relations (\ref{SOL3})--(\ref{SOL4B}) satisfy all three dynamic equations (\ref{A1})--(\ref{C1}).

Thus, the expressions (\ref{SOL3})--(\ref{SOL4B}) define the frame of exact solutions of cosmological dynamic equations (\ref{A1})--(\ref{C1}) for the inflationary models based on scalar-torsion gravity for arbitrary Hubble parameter $H=H(t)$.

Also, we note, that for the case teleparallel equivalent of general relativity (TERG) corresponding function is~\cite{Bahamonde:2021gfp}
\begin{eqnarray}
\label{TERG}
&& f(T,\phi)=-\frac{1}{2}T- V(\phi).
\end{eqnarray}

Since at the stage of cosmological inflation the rate of expansion of the universe is close to exponential regime (or close to the de Sitter stage) $T\simeq T_{0}=6H^{2}_{0}=const$, expanding $\sqrt{T}$ into a series around $T=T_{0}$ we get
\begin{eqnarray}
\label{DSAPPR}
&&\sqrt{T}=\frac{1}{2}\sqrt{T_{0}}+\frac{1}{2\sqrt{T_{0}}}T+\mathcal{O}\left[(T-T_{0})^{2}\right].
\end{eqnarray}

Thus, for quasi de Sitter inflationary stage from (\ref{SOL3}) and (\ref{DSAPPR}) we obtain
\begin{eqnarray}
\label{DSAPPR2}
&&f(T,\phi)\simeq\alpha_{1}\tilde{G}(\phi)T+\alpha_{2}\tilde{V}(\phi),
\end{eqnarray}
where $\tilde{G}(\phi)=\frac{G(\phi)}{2\sqrt{T_{0}}}$ and $\tilde{V}(\phi)=V(\phi)+T_{0}\tilde{G}(\phi)$.

Therefore, at the inflationary stage, exact expression for the function (\ref{SOL3}) can be reduced to the case of TERG (\ref{TERG}) for $\tilde{G}(\phi)=-1/(2\alpha_{1})=const$, $\alpha_{2}=-1$  with fairly high accuracy.

Now, we reconstruct the connection between the coupling function $G=G(\phi)$ and the scalar field potential $V=V(\phi)$ based on the nature of early universe accelerated expansion (close to exponential one) and observational constraints on the values of cosmological perturbations parameters.
For this purpose, let us consider cosmological perturbations in inflationary models based on the scalar-torsion gravity and generalized exact solutions (\ref{SOL3})--(\ref{SOL4B}) of cosmological dynamic equations (\ref{A1})--(\ref{C1}).

\section{Cosmological perturbations}\label{B}

In accordance with the theory of cosmological perturbations, quantum fluctuations of the scalar field induce the corresponding perturbations of the space-time metric during the inflationary stage. In the linear order of cosmological perturbation theory, the observed anisotropy and polarization of cosmic microwave background radiation (CMB)~\cite{Baumann:2014nda,Chervon:2019sey} are explained by the influence of two types of perturbations, namely, scalar and tensor ones.

Observational constraints on the parameters of cosmological perturbations due to the modern observations of the anisotropy and polarisation of CMB are~\cite{Planck:2018vyg,Galloni:2022mok}
\begin{eqnarray}
\label{PS}
&&P_S=2.1\times10^{-9},\\
\label{NS}
&&n_S=0.9649\pm 0.0042,\\
\label{R}
&&r<0.028.
\end{eqnarray}

The expressions for the parameters of cosmological perturbations at the crossing of the Hubble radius ($k=aH$) for the cosmological models based on action (\ref{action1}) can be written as follows~\cite{Gonzalez-Espinoza:2020azh}
\begin{eqnarray}
\label{pert1}
&&\mathcal{P}_S=\frac{H^2}{8\pi^2Q_{S}}\left[1+2\eta_{\mathcal{R}}\ln\left(\frac{k}{aH}\right)\right]_{k=aH}
=\frac{H^2}{8\pi^2Q_{S}},~~~~ \mathcal{P}_T=\frac{H^2}{2\pi^2Q_{T}},\\
\label{pert2}
&&n_S-1=-2\epsilon-\eta+2\eta_{\mathcal{R}},~~~~n_T=-2\epsilon-\delta_{f_{,T}},~~~~
r=\frac{\mathcal{P}_T}{\mathcal{P}_S}=16\delta_{\omega X},
\end{eqnarray}
where $\mathcal{P}_S$ and $\mathcal{P}_T$ are the power spectra of scalar and tensor perturbations, $n_{S}$ and $n_T$ are the spectral index of the scalar and tensor perturbations and $r$ is the tensor-to-scalar ratio. Also, $n_{T}<0$ corresponds to the red tilt tensor spectrum, $n_{T}>0$ corresponds to the blue tilt tensor spectrum and $n_{T}=0$ corresponds to the flat tensor spectrum.

Also, the parameters $Q_{S}$ and $Q_{T}$ are~\cite{Gonzalez-Espinoza:2020azh}
\begin{eqnarray}
\label{pert2A}
&&Q_{S}=\frac{\omega X}{H^{2}},~~~Q_{T}=-\frac{1}{2}f_{,T},
\end{eqnarray}
and the slow-roll parameters in expressions (\ref{pert1})--(\ref{pert2}) are defined as follows
\begin{eqnarray}
\label{pert4}
&&\epsilon=-\frac{\dot{H}}{H^2},~~~~\delta_{\omega X}=-\frac{\omega X}{2H^2f_{,T}},~~~~ \delta_{f_{,T}}=\frac{\dot{f}_{,T}}{Hf_{,T}},\\
\label{pert5}
&&\delta_{f\dot{H}}=\frac{f_{,TT}\dot{T}}{Hf_{,T}},~~~\delta_{fX}=\frac{f_{,T\phi}\dot{\phi}}{Hf_{,T}},~~~
\eta_{\mathcal{R}}=\delta_{f_{,T}}\left[1+\frac{\delta_{f_{,T}}}
{\delta_{f\dot{H}}}\left(1+\frac{\delta_{fX}}{\delta_{\omega X}}   \right)\right],\\
\label{pert6}
&&\eta=\frac{\dot{Q}_{S}}{HQ_{S}}=3\epsilon+\frac{1}{H}\left[\frac{d}{dt}\ln(\omega X H)\right]=
2\epsilon+\frac{1}{H}\left(\frac{\dot{\omega}}{\omega}+\frac{\dot{X}}{X}\right),
\end{eqnarray}
where all these parameters much less than unity due to the quasi de Sitter inflationary regime of the early universe's dynamics~\cite{Gonzalez-Espinoza:2019ajd,Gonzalez-Espinoza:2020azh}.

On the basis of expressions (\ref{SOL4})--(\ref{SOL4B}) for the proposed exact solutions we obtain following expression
\begin{eqnarray}
\label{EXPRESSION}
\omega X=\frac{\alpha_{1}}{\sqrt{6}}\dot{G}=\alpha_{2}V.
\end{eqnarray}

Thus, from (\ref{PS}), (\ref{pert1}), (\ref{pert2A}) and  (\ref{EXPRESSION}) we obtain following condition
\begin{eqnarray}
\label{SRD3}
&&\mathcal{P}_S=\left(\frac{H^2}{8\pi^2Q_{S}}\right)_{k=aH}=
\left(\frac{H^4}{8\pi^2\alpha_{2}V}\right)_{k=aH}
=2.1\times10^{-9}.
\end{eqnarray}

Since condition (\ref{SRD3}) can always be satisfied by choosing the parameters of the inflationary model, as a criterion for verifying cosmological models we will consider the dependence of the tensor-scalar ratio on the spectral index of scalar perturbations $r=r(1-n_{S})$.

Taking into account (\ref{EXPRESSION}) and expression $T=6H^{2}$, for the inflationary models described by solutions (\ref{SOL3})--(\ref{SOL4B}) from expressions (\ref{pert4})--(\ref{pert6}) we obtain
\begin{eqnarray}
\label{pert7}
&&\delta_{\omega X}=\frac{\sqrt{6}\,\omega X}{HG}=\frac{\alpha_{1}\dot{G}}{HG},~~~~\delta_{f_{,T}}=\epsilon+\delta_{fX},~~~~
\delta_{f\dot{H}}=\epsilon,\\
\label{pert8}
&&\delta_{fX}=\frac{\sqrt{6}\alpha_{2}}{\alpha_{1}}\,\frac{V}{HG}=
\frac{\dot{G}}{HG}=\frac{\delta_{\omega X}}{\alpha_{1}},~~~~~~~
\eta=2\epsilon+\frac{\ddot{G}}{H\dot{G}},\\
\label{pert9}
\nonumber
&&\eta_{\mathcal{R}}=\delta_{f_{,T}}\left[1+\left(1+\frac{\delta_{fX}}{\epsilon}\right)
\left(1+\frac{1}{\alpha_{1}}\right)\right]=\\
&&=\left(\epsilon+\frac{\dot{G}}{HG}\right)\left[1+\left(1-\frac{H\dot{G}}{G\dot{H}}\right)
\left(1+\frac{1}{\alpha_{1}}\right)\right].
\end{eqnarray}

Thus, we can write the spectral indices of the scalar and tensor perturbations and the tensor-to-scalar ratio at the crossing of the Hubble radius as
\begin{eqnarray}
\label{pert11}
&&n_{S}-1=4\frac{\dot{H}}{H^{2}}-\frac{\ddot{G}}{H\dot{G}}+
2\left(-\frac{\dot{H}}{H^{2}}+\frac{\dot{G}}{HG}\right)\left[1+\left(1-\frac{H\dot{G}}{G\dot{H}}\right)
\left(1+\frac{1}{\alpha_{1}}\right)\right],\\
\label{pert12}
&&r=16\alpha_{1}\frac{\dot{G}}{HG},\\
\label{pert11T}
&&n_{T}=3\frac{\dot{H}}{H^{2}}-\frac{\dot{G}}{HG}=-3\epsilon-\frac{r}{16\alpha_{1}}.
\end{eqnarray}

As one can see from expression (\ref{pert11T}), for $\alpha_{1}>-\frac{r}{48\epsilon}$ one has red tilt tensor spectrum, for $\alpha_{1}<-\frac{r}{48\epsilon}$ one has blue tilt tensor spectrum, and for $\alpha_{1}=-\frac{r}{48\epsilon}$ one has flat tensor spectrum.

\section{Reconstruction of the parameters of verified inflationary models  }\label{C}

Now, we will determine the connection between the coupling function $G$ and the Hubble parameter $H$, for dependence $r=r(1-n_{S})$ in the main (linear) order and consider quasi de Sitter solutions for the model under consideration.

Since the value of the spectral index of scalar perturbations is $n_{S}\simeq0.97$ and $1-n_{S}\simeq0.03\ll1$, we can write the dependence $r=r(1-n_{S})$ as follows
\begin{eqnarray}
\label{pert13}
&&r=\sum^{\infty}_{k=0}\beta_{k}(1-n_{S})^{k}=\beta_{0}+\beta_{1}(1-n_{S})+\beta_{2}(1-n_{S})^{2}+...,
\end{eqnarray}
where $(1-n_{S})$ is the small parameter of expansion and $\beta_{k}$ are the constant coefficients.

Since, the zeroth order term in this expansion $r(0)=\beta_{0}=0$ from condition $r(n_{S}=1)=0$ corresponding to the flat Harrison-Zel'dovich spectrum \cite{Baumann:2014nda,Chervon:2019sey}, we can rewrite expression (\ref{pert13}) in the following form
\begin{eqnarray}
\label{RNSEXPANSION}
&&r=\sum^{\infty}_{k=1}\beta_{k}(1-n_{S})^{k}=\beta_{1}(1-n_{S})+\beta_{2}(1-n_{S})^{2}+...
\end{eqnarray}

Thus, we can consider the new specific classification of cosmological models (regardless of gravity type or cosmological model's parameters) according to the orders of expansion of dependence $r=r(1-n_{S})$.

Since the first term in expression (\ref{RNSEXPANSION}) makes the main contribution to the value of the tensor-to-scalar ratio, we can consider the dependence $r=r(1-n_{S})$ with a reasonable degree of accuracy in the first order $r\sim(1-n_{S})$ on the basis of expressions (\ref{pert11})--(\ref{pert12}).

For the purpose to obtain the connection between parameters of the cosmological models under consideration for the first-order models with $r\sim(1-n_{S})$ in explicit form with arbitrary Hubble parameter (i.e. for arbitrary cosmological dynamics), we consider $\alpha_{1}=-1$. For this case, from expressions (\ref{pert11})--(\ref{pert11T}) we obtain
\begin{eqnarray}
\label{pert11L}
&&n_{S}-1=2\frac{\dot{H}}{H^{2}}+2\frac{\dot{G}}{HG}-\frac{\ddot{G}}{H\dot{G}},\\
\label{pert12L}
&&r=-16\frac{\dot{G}}{HG},\\
\label{pert13T}
&&n_{T}=-3\epsilon+\frac{r}{16}.
\end{eqnarray}

Also, we note, that for the partial case $G\sim H$, from (\ref{pert11L})--(\ref{pert13T}) we obtain
\begin{eqnarray}
\label{TERGpert}
&&n_{S}-1=-4\epsilon+2\delta,~~~n_{T}=-2\epsilon,~~~~r=16\epsilon,
\end{eqnarray}
where $\delta=-\frac{\ddot{H}} {2H\dot{H}}$, that corresponds to the case of TEGR~\cite{Gonzalez-Espinoza:2019ajd} for the scalar-torsion gravity $f(T,\phi)_{STG}=- G(\phi)\sqrt{T}$ under consideration. In such a case, inflationary models with only some certain potentials of the scalar field will satisfy the observational constraints on the parameters of cosmological perturbations~\cite{Chervon:2023gio}.

In general case, the linear dependence between tensor-to-scalar ratio and spectral index of scalar perturbations can be defined as follows
\begin{eqnarray}
\label{RNS}
&&r=\frac{8}{m}(1-n_{S}),
\end{eqnarray}
where $\beta_{1}=8/m$ and $m>0$ is the positive constant.

Also, on the basis of the observational constraints (\ref{NS})--(\ref{R}) from (\ref{RNS}) for verifiable inflationary models one has following constraint $m>11$. We also note that any future constraints on the values of the cosmological perturbation parameters only lead to a change in the estimate of the constant parameter $m$. Thus, models of cosmological inflation based on relation (\ref{RNS}) are fundamentally verified by observational constraints following from observations of the anisotropy and polarization of CMB~\cite{Planck:2018vyg,Galloni:2022mok}.

From equations (\ref{pert11L})--(\ref{pert12L}) and (\ref{RNS}) we can reconstruct corresponding connection between Hubble parameter $H$ and coupling function $G$ for this case, namely
\begin{eqnarray}
\label{RNSH}
&&H(t)=const\times\dot{G}^{1/2} G^{m-1}.
\end{eqnarray}

From equations (\ref{SOL4B}) and (\ref{RNSH}) for such a models we obtain
\begin{eqnarray}
\label{RNSHV}
&&H(t)=const\times V^{1/2}G^{m-1},
\end{eqnarray}
where the Hubble parameter $H>0$ and has the real values for any type of potential (positive or negative) due to the fact that the constant in expression (\ref{RNSHV}) is arbitrary one.

For the pure de Sitter regime of accelerated expansion of the early universe $H=const$ from (\ref{RNSHV}) we obtain
\begin{eqnarray}
\label{SRD1}
&&G(\phi)= const\times\left[V(\phi)\right]^{-\frac{1}{2(m-1)}},
\end{eqnarray}
for any type of the scalar field evolution $\phi=\phi(t)$.

Therefore, taking into account quasi de Sitter regime of accelerated expansion of the early universe $H\simeq const$, we can represent the  expression for the coupling function as follows
\begin{eqnarray}
\label{CF1}
&&G(\phi)\sim\left[V(\phi)\right]^{-\frac{1+k}{2(m-1)}},
\end{eqnarray}
where parameter $|k|\ll1$ defines the deviations from the pure de Sitter stage.

Therefore, the gravitational term in action (\ref{action1}) for exact relations (\ref{SOL3})--(\ref{SOL4B}) in quasi de Sitter regime $H\simeq const$ can be defined as
\begin{eqnarray}
\label{SRD2}
&&f(T,\phi)_{STG}=- G(\phi)\sqrt{T}\sim\sqrt{T}\times\left[V(\phi)\right]^{-\frac{1+k}{2(m-1)}}.
\end{eqnarray}

Thus, for any physical potential of a scalar field~\cite{Martin:2013tda}, from expression (\ref{SRD2}) one can reconstruct the type of scalar-torsion gravity corresponding to the possibility of verifying different realizations of the inflationary scenario.

\section{Conclusion}

In this work, cosmological models based on the certain type of the scalar-torsion gravity $f(T,\phi)_{STG}\sim G(\phi)\sqrt{T}$ are considered. This type of the scalar-torsion gravity was obtained as consequence of the generalized exact solutions of cosmological dynamic equations.

A classification of inflation models according to the order of expansion of the dependence tensor-to-scalar ratio from spectral index of the scalar perturbations $r=r(1-n_{S})$ was also proposed. On the basis of this classification the method for constructing verified inflationary models based on scalar-torsion gravity $f(T,\phi)_{STG}=-G(\phi)\sqrt{T}$ for the first-order models $r\sim(1-n_{S})$ was considered.
In this case, the type of scalar field potential $V=V(\phi)$ or other background parameters don't affect the possibility of verifying the proposed cosmological models.

Also, due to the correspondence of arbitrary models of cosmological inflation based on the generalized exact solutions to observational constraints on the parameters of cosmological perturbations, the following relationship between the non-minimal coupling function and the potential of the scalar field $G(\phi)\sim\left[V(\phi)\right]^{-\frac{1+k}{2(m-1)}}$ for quasi de Sitter dynamics accelerated expansion of the universe was obtained. Thus, the type of non-minimal coupling between the scalar field and torsion depends on the type of the scalar field potential for the proposed inflationary models.

In conclusion, we note that proposed type of the scalar-torsion gravity $f(T,\phi)_{STG}\sim G(\phi)\sqrt{T}$ reconstructed from exact solutions of cosmological dynamic equations and corresponding wide class of verified cosmological models with arbitrary parameters is of interest for the further research of the deviations in the spectrum of relict gravitational waves from teleparallel equivalent of general relativity or from the other modified gravity theories.


%
%

%
%

\end{document}